\documentstyle[preprint,aps]{revtex}
\newcommand\DD {$\Delta {\mathrm (1232)}$ }
\newcommand\D {$\Delta$ }
\newcommand\nini {Ni + Ni }
\begin{document}
\preprint{ }
\draft

\title{
Abundance of $\Delta$ Resonances in $^{58}$Ni + $^{58}$Ni Collisions\\
between 1 and 2 AGeV
}

\author{
B.~Hong$^{4}$\footnote{Present address: 
Korea University, Seoul, South Korea}, 
N.~Herrmann$^{4,6}$, J. L.~Ritman$^{4}$, D.~Best$^{4}$, 
A.~Gobbi$^{4}$, K.~D.~Hildenbrand$^{4}$, M.~Kirejczyk$^{4,10}$, 
Y.~Leifels$^{4}$, C.~Pinkenburg$^{4}$, W.~Reisdorf$^{4}$, D.~Sch\"ull$^{4}$, 
U.~Sodan$^{4}$, G.~S.~Wang$^{4}$, T.~Wienold$^{4}$, J.~P.~Alard$^{3}$, 
V.~Amouroux$^{3}$, 
N.~Bastid$^{3}$, I.~Belyaev$^{7}$, G.~Berek$^{2}$, J.~Biegansky$^{5}$, 
A.~Buta$^{1}$, 
J.~P.~Coffin$^{9}$, P.~Crochet$^{9}$, R.~Dona$^{9}$, P.~Dupieux$^{3}$, 
M.~Eskef$^{6}$, P.~Fintz$^{9}$, Z.~Fodor$^{2}$, L. Fraysse$^{3}$,
A.~Genoux-Lubain$^{3}$, G.~Goebels$^{6}$, G.~Guillaume$^{9}$, 
E.~H\"afele$^{6}$, 
F.~Jundt$^{9}$, J.~Kecskemeti$^{2}$, M.~Korolija$^{6}$, R.~Kotte$^{5}$, 
C.~Kuhn$^{9}$, A.~Lebedev$^{7}$, I.~Legrand$^{1}$, C.~Maazouzi$^{9}$, 
V.~Manko$^{8}$, J.~M\"osner$^{5}$, S.~Mohren$^{6}$, W.~Neubert$^{5}$, 
D.~Pelte$^{6}$, M.~Petrovici$^{1}$, P. Pras$^{3}$, F.~Rami$^{9}$, 
C.~Roy$^{9}$, Z.~Seres$^{2}$, B.~Sikora$^{10}$, V.~Simion$^{1}$, 
K.~Siwek-Wilczy\'{n}ska$^{10}$, A.~Somov$^{7}$, L.~Tizniti$^{9}$, 
M.~Trzaska$^{6}$, M.~A.~Vasiliev$^{8}$, P.~Wagner$^{9}$, 
D.~Wohlfarth$^{5}$, A.~Zhilin$^{7}$\\
(FOPI Collaboration)
}

\address{
$^1$ Institute for Nuclear Physics and Engineering, Bucharest, Romania\\
$^2$ Central Research Institute for Physics, Budapest, Hungary\\
$^3$ Laboratoire de Physique Corpusculaire, IN2P3/CNRS, and Universit\'{e}
Blaise Pascal, Clermont-Ferrand, France\\
$^4$ Gesellschaft f\"ur Schwerionenforschung, Darmstadt, Germany\\
$^5$ Forschungszentrum Rossendorf, Dresden, Germany\\
$^6$ Physikalisches Institut der Universit\"at Heidelberg, Heidelberg, Germany\\
$^7$ Institute for Theoretical and Experimental Physics, Moscow, Russia\\
$^8$ Kurchatov Institute, Moscow, Russia\\
$^9$ Centre de Recherches Nucl\'{e}aires and Universit\'{e} Louis Pasteur,
Strasbourg, France\\
$^{10}$ Institute of Experimental Physics, Warsaw University, Poland\\
}

\maketitle

\begin{abstract}
Charged pion spectra measured in $^{58}$Ni-$^{58}$Ni collisions
at 1.06, 1.45 and 1.93 AGeV
are interpreted in terms of a thermal model including the decay of 
\D resonances.
The transverse momentum spectra of pions are well reproduced by adding 
the pions originating from the \D-resonance decay to the 
component of thermal pions, deduced from the high 
transverse momentum part of the pion spectra. About 10 and 18$\%$ 
of the nucleons are excited to \D states at freeze-out 
for beam energies of 1 and 2 AGeV, 
respectively.
\end{abstract}
\vspace{1.0cm}


Relativistic heavy ion collisions are a unique tool to investigate 
hot and dense nuclear matter in the laboratory\cite{rs1,sto1}. 
In these studies the measurement of pions 
has attracted much interest since they can be regarded
as a potential probe of the processes 
beyond mere nucleon-nucleon binary interactions in heavy ion 
collisions\cite{sto2,pda1,sto3,hari1,bass2}. According to the results 
of microscopic model calculations\cite{sto1}, pions are, in principle, 
expected to contain information from the early stage of the fireball 
to the freeze-out, since they are produced continuously during the whole 
reaction process. However, the large pion-nucleon cross section destroys 
the memory to early times, and the signatures of the reaction dynamics have 
disappeared by the time of detection. Therefore one will, most likely, 
observe only the conditions at freeze-out.
On the other hand, this knowledge is important in itself, 
because it gives information on the degree of equilibration. 
To a large extent, a model that assumes thermal and hadrochemical 
equilibrium can explain the various particle yield ratios including 
multistrange particles at AGS and SPS energies\cite{pbm1}.

The possibility of exciting \DD resonances (called \D in the following)
and hence to form so-called \D matter at incident energies
between 1 and 10 AGeV has been investigated theoretically in 
RQMD simulations\cite{mh1}.  
The model predicts about 10 $\%$ of the participant nucleons being excited
to \D during the hot and dense phase  
at 1 AGeV, while at 10 AGeV more than 50 $\%$ of the 
nucleons are excited to baryonic resonances, of which only about half 
are present as \D. Pions from the resonance decays show a 
different phase space distribution depending on their decay 
momentum\cite{sol1}, and especially the \D\,-\,decay kinamatics has been 
made responsible for the low 
transverse momentum ($p_{t}$) enhancement at incident beam energies from 
1 to 15 AGeV\cite{geb1,wein1,e814-1}. The concave shape found 
in the Boltzmann (or invariant) spectra of pions at beam energies from 
0.5 to 200 AGeV\cite{e814-1,rbr1,bh1,cmuz1,na35-1,na44-1}
confirms the importance of the resonance decay. 
In this paper we will investigate the \D  population at freeze-out,
using the spectral shape of pion distributions obtained in the Ni+Ni system 
at incident energies varying from 1 to 2 AGeV.

The experiment has been performed with the FOPI detection system\cite{ag1,jr1}
at the heavy ion synchrotron SIS at GSI by bombarding
a $^{58}$Ni target of 225 mg/cm$^{2}$ thickness
($\approx$ 1 $\%$ nuclear interaction length) 
with $^{58}$Ni beams of 1.06, 1.45 and 1.93 AGeV.  
For the analysis presented here the central drift chamber (CDC),
working in the 0.6~T field of a solenoidal magnet has been used 
for particle identification, and the forward plastic wall for the
centrality determination. 
This azimuthally symmetric wall 
covers polar angles $\theta_{Lab}$ from 7 to 30 degrees, measuring energy
loss $dE/dx$ and time of flight of the charged fragments
the multiplicity of which  
(PMUL) is used for centrality selection. The CDC covers 
$\theta_{Lab}$ from 30 to 150 degrees; identification of the pions and other
light charged fragments is achieved via 
$dE/dx$ and momentum measurement. 
To compare the rapidity distributions at different beam energies 
we define the normalized rapidity $y^{(0)} \equiv y / y_{cm} - 1$,
where $y_{cm}$ is the rapidity of the center of mass system (c.m.). 
Throughout this paper we use 
the convention $\hbar$ = c = 1.

\begin{figure}
\caption{ 
Boltzmann spectra of $\pi^{-}$ for different windows in normalized rapidity
(\nini collisions at 1.93 AGeV under selectivity cut PM100, see
text for definition); the two upper curves have been multiplied by the
factors 10$^{2}$ and 10$^{4}$, respectively. 
The black solid lines represent the model calculations including the
thermal pions (dashed) as well as pions from the \D decay (dotted).
Pions from the decay of higher baryonic resonances, e.g. from
$N^{\star}$(1440) (dash-dotted), $N^{\star}$(1520) (grey solid-a)
and $N^{\star}$(1535) (grey solid-b) decays, are shown as well.
}
\label{pibolz}

\vspace{16cm}
\includegraphics{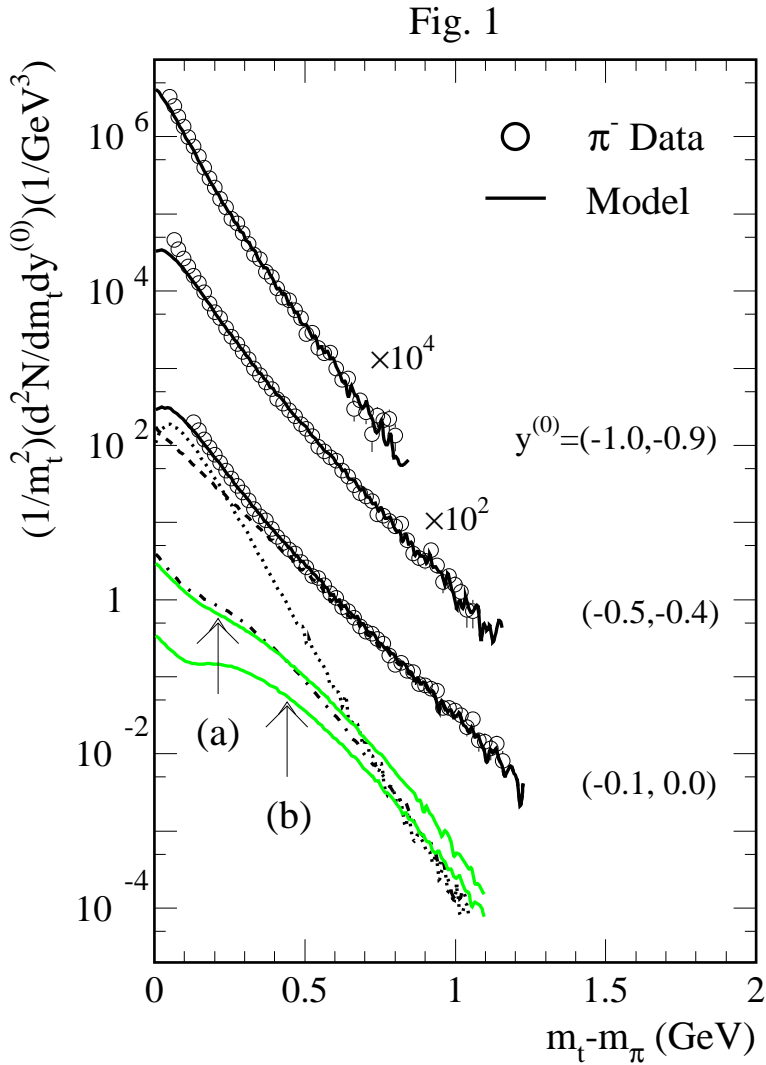}
\end{figure}

We have previously observed that the Boltzmann spectra,
$d^{2} N/m_{t}^{2} dm_{t} dy^{(0)}$ with 
$m_{t}=\sqrt{p_{t}^{2} + m_{\pi}^{2}}$ being the transverse mass 
and $m_{\pi}$ the pion mass, of $\pi^{\pm}$ measured in reactions
between 1 and 2 AGeV show a 
concave shape which can be fitted by the sum of two exponential 
functions with a $\chi^{2}$ per degree of freedom $\sim$ 1 in all rapidity 
bins\cite{bh1}. Fig.~\ref{pibolz} shows, as example, the Boltzmann spectra 
of $\pi^{-}$ at 1.93 AGeV with a cut in PMUL selecting the uppermost 100 mb 
(or 4$\%$ of the total reaction cross section, called PM100 in the following) 
for different rapidity windows. Only the statistical errors 
are given in the spectra; the systematic error in the absolute 
normalization~\cite{bh1}, estimated to
about 10 $\%$, is not shown. 
It has been demonstrated that at freeze-out the \D-decay pions 
contribute dominantly to the low kinetic energy part, whereas 
thermal pions contribute predominantly to the high kinetic 
energy part of the spectra both at SIS and AGS  
energies\cite{geb1,wein1,e814-1}. Therefore we attribute the high-$p_{t}$ 
component of the experimental spectra as being due to thermal pions. 

In order to decompose the pion spectra quantitatively we 
calculate the part arising from \D decay by a Monte Carlo routine.
For the phase space distribution of the \D at freeze-out, we use the 
isotropically expanding blast model proposed by Siemens 
and Rasmussen\cite{psie1}:
\begin{equation}
{{d^{2} N_{\Delta}} \over {p_{t} d p_{t} dy}} \propto
\int_{m_{N}+m_{\pi}}^{\infty} dm_{\Delta}
P_{\Delta}(m_{\Delta}) E_{\Delta} e^{-\gamma_{r} E_{\Delta} / T}
[(\gamma_{r} + {{T}\over{E_{\Delta}}}){{\sinh \alpha} \over {\alpha}} -
{{T} \over {E_{\Delta}}}\cosh \alpha],
\label{shellflow}
\end{equation}
where $m_{N}$ is the nucleon mass, 
$E_{\Delta} = \sqrt{p_{\Delta}^{2}+m_{\Delta}^{2}}$ and $p_{\Delta}$ are 
the total energy and momentum of the \D in the c.m.,
$\gamma_{r} = 1 / \sqrt{1-\beta_{r}^{2}}$, and
$\alpha = (\gamma_{r} \beta_{r} p_{\Delta}) / T$.
For the freeze-out temperature ($T$) and radial flow velocity 
($\beta_{r}$), the results from Ref.\cite{bh1} are used, e.g.
$T =$ 92 MeV and $\beta_{r} =$ 0.32 at 1.93 AGeV,
obtained in analyses of the high-$p_{t}$ part of pion, 
proton and deuteron spectra at midrapidity. 
We adopt an isotropically emitting source scenario that 
describes the high $p_t$ part of the pion spectrum very well. 
It should be noted, however, that the proton rapidity 
distributions are more elongated and the apparent temperatures near
target rapidity are smaller as compared to the expectations from 
an isotropically radiating source \cite{bh1}. 
Most probably this is caused by the interaction of participating nucleons 
with some spectator remnants or partial transparency of the system.
For those scenarios a reliable prescription for the \D phase space 
distribution does not exist at the moment.
The \D  mass distribution, 
$P_{\Delta}(m_{\Delta})$, is parameterized by a 
relativistic Breit-Wigner form \cite{gino1} as follows:
\begin{equation}
P_{\Delta} (m_{\Delta}) \propto ({1 \over {q^{2}}})
{{\Gamma_{\Delta}^{2}(q)} \over {(m_{\Delta}^{2} - \bar{m}_{\Delta}^{2})^{2} 
+ \bar{m}_{\Delta}^{2} {\Gamma_{\Delta}^{2}(q)}}},
\label{prob}
\end{equation}
where 
\begin{equation}
\Gamma_{\Delta} (q) = \bar{\Gamma}_{\Delta}
({q \over {\bar{q}}})^{3}
{{1 + (R_{1} \bar{q})^{2} + (R_{2} \bar{q})^{4}} \over
{1 + (R_{1} q)^{2} + (R_{2} q)^{4}}}
\label{qwidth}
\end{equation}
is the momentum dependent width of the \D, $q$ is the momentum of the
pion in the \D  rest frame. We chose $\bar{\Gamma}_{\Delta} =$ 112 MeV,
$\bar{q} =$ 227 MeV, $R_{1} =$ 0.0042 MeV$^{-1}$ and $R_{2} =$ 
0.0032 MeV$^{-1}$. The maximum of $P_{\Delta} (m_{\Delta})$ is fixed to 
the free \D mass, i.e. $\bar{m}_{\Delta} =$ 1232 MeV. 

After superposition with the spectra of thermal pions
the results of the model calculation are shown by the black solid lines 
in Fig.~\ref{pibolz}. We use for all rapidity bins only 
one normalization factor for
the relative abundance of the decay pions with respect to thermal 
pions. The overall agreement between 
the model and data is rather good except at very low $m_{t}$ 
($m_{t} - m_{\pi} \leq$ 0.12 GeV), where the data exceed the model 
calculations at all rapidities. In order to explain this difference, 
one may need to consider the decrease of the effective \D mass as shown, 
e.g. by the thermal model including the $\pi-$nucleon loop correction 
to the \D self-energy calculation\cite{wein1}.
The decomposition 
of $\pi^{-}$ spectra into $\pi_{T}$ (dashed line) 
and $\pi_{\Delta}$ (dotted line) at midrapidity is also displayed 
in Fig.~\ref{pibolz}, demonstrating that the low-$p_{t}$ enhancement 
of the pion spectra is consistent with the contribution from \D  decay. 
The influence of the $p_{t}$ resolution ($\sigma_{p_{t}} / p_{t}$)
on the  $\pi_{\Delta}$ spectra 
which worsens from 4 $\%$ at $p_{t} \leq$ 0.5 GeV to about 9 $\%$ 
at $p_{t} =$ 1.5 GeV\cite{dp1} has been 
tested, too: a slight effect can be found only for $m_{t} - m_{\pi} \geq$ 1 GeV,
and it does not affect our results.

The influence of higher-mass baryonic resonances (e.g. N$^{\star}$(1440),
N$^{\star}$(1520) and N$^{\star}$(1535)) has been studied using the 
hadrochemical equilibrium model\cite{pbm1,bh2}. The model parameters, 
chemical freeze-out temperature $T_{C}$ and baryon chemical potential 
$\mu_{B}$, are adjusted to reproduce the measured particle yields of 
high-$p_{t}$ (thermal) pions, protons, deuterons and $\eta '$s\cite{bh1,taps}. 
The temperatures $T_{C}$ from the hadrochemical equilibrium model agree 
with those from the radial flow analysis, $T$, within 8 $\%$~\cite{bh2}. The
Boltzmann spectra of pions from the N$^{\star}$ resonance decay at midrapidity 
are also shown in Fig.~\ref{pibolz}, calculated with the proper 
branching ratios under the assumption of isotropic
expanding thermal distributions 
of primary resonances (Eq.~\ref{shellflow}) with a fixed mass. 
The total contribution of N$^{\star}$ resonances to our estimation of the 
\D abundance is at most 5$\%$, in agreement with earlier RQMD 
and BUU studies\cite{mh1,bali1}; hence this contribution
is neglected in the current analysis.

The concave shape in the pion spectra can also be found in 
transport models\cite{bali1,bass1}, although the high-$p_{t}$ pions are of
different origin there. Within these models, all pions stem from the decay of 
resonances. Pions in the high-$p_{t}$ tail have decoupled earlier
from the system ($free$ pions), when more energetic resonances were populated.
The pion spectra are obtained by integrating over the complete compression
expansion cycle. 
However, when analysing the result of the reaction under the assumption 
of a sudden freeze-out, the spectrum of the $free$ pions 
exhibits an exponential 
shape as obtained for the thermal pions in the present equilibrium model.

\begin{figure}
\caption{
Top: Rapidity spectra of $\pi^{-}$ at 1.06, 1.45 and 1.93 AGeV under 
the centrality cut PM100. 
The solid circles represent the integrations
from $p_{t} =$ 0 to $\infty$ of the fit 
with two exponentials ($\pi^{-}_{All}$), while the open circles 
represent the integrations of the low-$p_{t}$ exponential fit only 
($\pi^{-}_{Low}$). Bottom: Ratio of $\pi^{-}_{Low} / \pi^{-}_{All}$; 
the dashed lines represent the mean values of the ratio for all rapidities.
}
\label{dndy}

\vspace{15cm}
\includegraphics{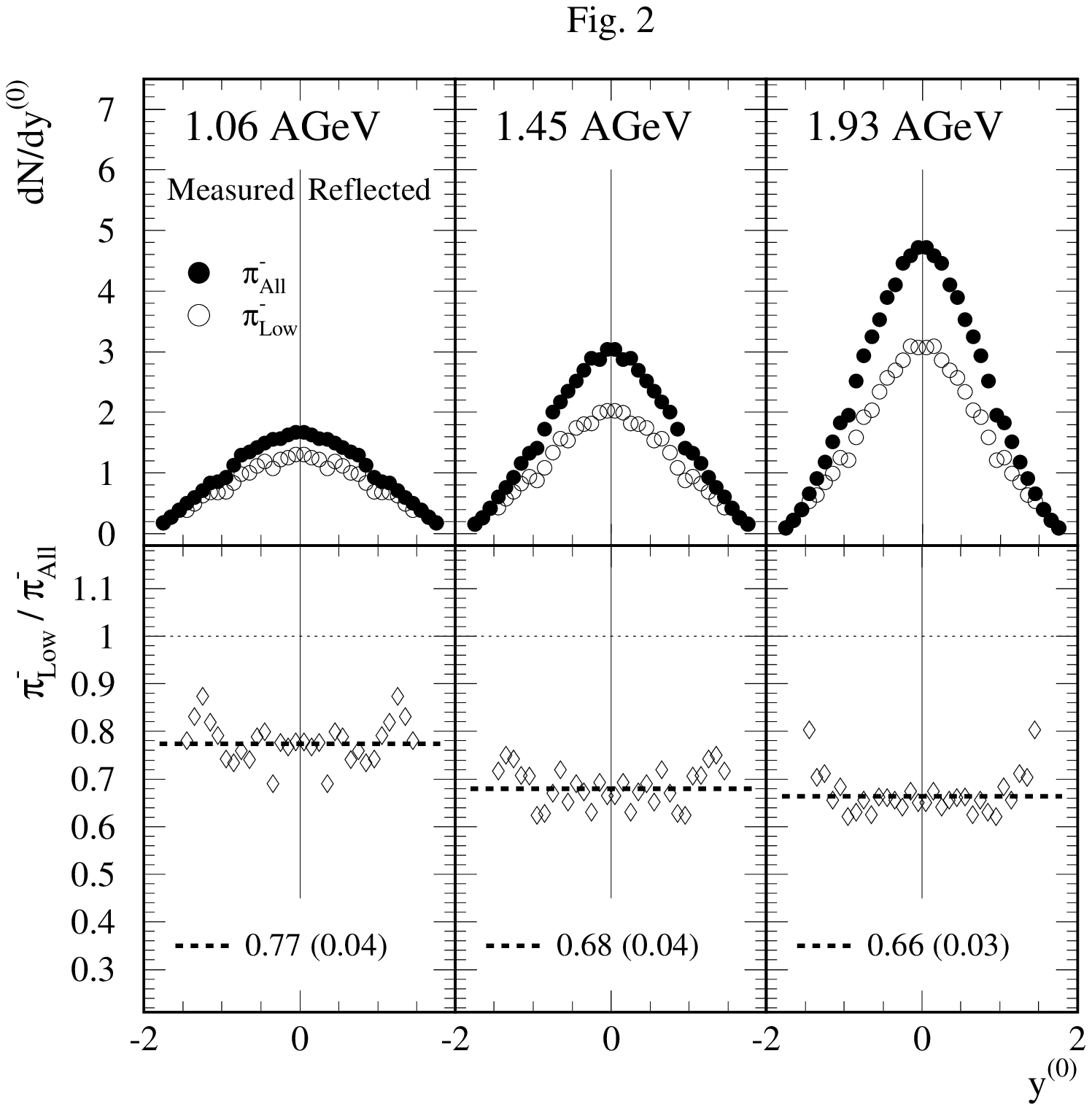}
\end{figure}

To estimate the yields of all pions ($\pi^{-}_{All}$) and that of the  
low-$p_{t}$ pions ($\pi^{-}_{Low}$) 
we have integrated the fitted function
(sum of two exponentials) over the whole range
$p_{t} =$ 0 to $\infty$ and for the
low-$p_{t}$ exponential only, respectively. 
Since the slope parameter of the low $p_t$ component from the two 
exponential fit agree with those of the pion spectra from \D decay 
within 6\% for all beam energies,
we simply take $\pi_{low}$ multiplied with the proper isobar ratio 
as measure of the total number of \D in the system.
The resulting $dN/dy^{(0)}$
spectra are shown in the top panel of Fig.~\ref{dndy} for the three 
beam energies under the PM100 cut. 
Note that only data for $y^{(0)} <$ 0 are measured, 
and then reflected to $y^{(0)} >$ 0, employing the symmetry of the 
colliding system. 
The total $\pi^{-}$ yield per event, $n(\pi^{-})$, is summarized
in the second column of Table~\ref{abund}. The number of participant 
nucleons ($A_{part}$) is calculated via the geometrical model\cite{dp1}; 
for PM100  events in \nini collisions one obtains 
101 $\pm$ 8 nucleons in the central fireball. In comparison to the earlier 
BEVALAC data for Ar + KCl\cite{hari1}, 
our measurement of the $\pi^{-}$ multiplicity per participant nucleon, 
$n(\pi^{-}) / A_{part}$, is about 15 $\%$ lower than the mean value
given there; 
within the error bars both measurements are compatible. The ratio 
$\pi^{-}_{Low} / \pi^{-}_{All}$ is shown in the bottom panel of 
Fig.~\ref{dndy}, where the dashed lines represent the mean of the ratio 
for the whole rapidity bins given also in the third column of table I. 
The systematic error of  $\pi^{-}_{Low} / \pi^{-}_{All}$ introduced 
by using the exponential fit at low $p_t$ rather than the 
$\pi_{\Delta}$ spectrum is estimated to less than 2\%.
The ratio
decreases from about 77 $\%$ to 66 $\%$ as we increase the beam energy 
from 1.06 to 1.93 AGeV. At 14.6 AGeV this $\pi^{-}_{Low} / \pi^{-}_{All}$ 
ratio was determined to 33 $\pm$ 5 $\%$ \cite{e814-1}.

\begin{figure}
\caption{
\label{delapart}
Ratio $n(\Delta) / n({\mathrm{nucleon}}+\Delta)$ at freeze-out as 
a function of the beam energy.} The open square is the result of the  
analysis of experiment E814 at 14.6 AGeV \cite{e814-1}. 

\vspace{15cm}
\includegraphics{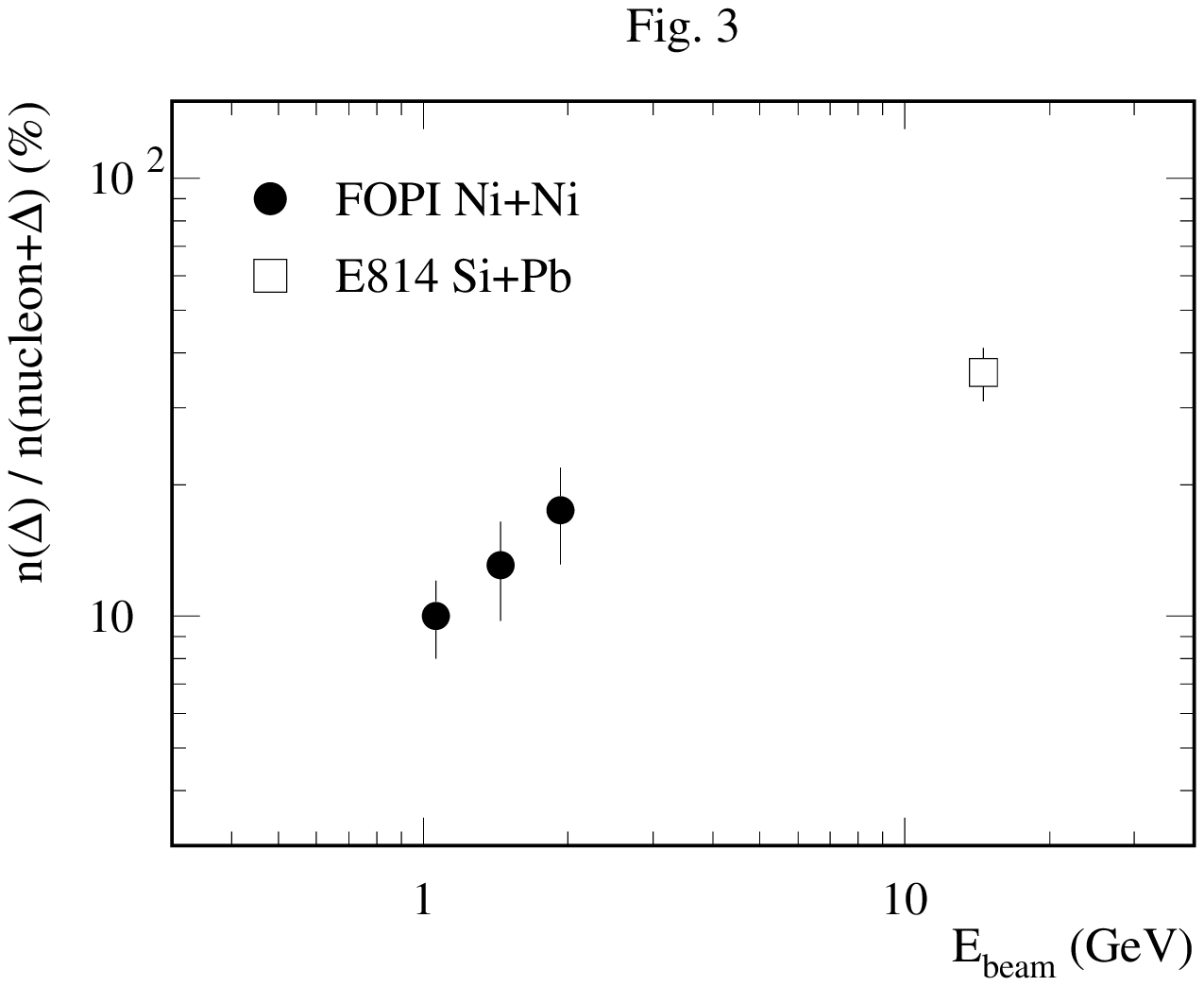}
\end{figure}

Having this information, we can now estimate the \D abundance  
$n(\Delta)$ at freeze-out by means of the following relation:
\begin{equation}
n(\Delta) = n(\pi^{-}) \times f_{isobar} \times 
({{\pi^{-}_{Low}} \over {\pi^{-}_{All}}}),
\label{delest}
\end{equation}
where 
\begin{equation}
f_{isobar} \equiv {{n(\pi^{+} + \pi^{0} + \pi^{-})} \over {n(\pi^{-})}} 
= {{6~(Z^{2} + N^{2} + NZ)} \over {5~N^{2} + NZ}}
\label{isobar}
\end{equation}
is the prediction of the isobar model\cite{rs1}
($N$ is the number of neutrons, $Z$ the number of protons).
For the energy range considered here, this model describes the relative 
abundance of pions in central collisions within 10\% \cite{dp1,dp2}.
For $^{58}$Ni 
one obtains $f_{isobar}$=2.84. 
Note that for the present system $N$ is close to $Z$,
hence one does not expect big deviations from $f_{isobar}$=3 in any model. 
Eq.~\ref{delest} contains the implicit
assumption that the
ratio $\pi_{Low} / \pi_{All}$ is the same for $\pi^{+}$, $\pi^{0}$
and $\pi^{-}$. From our analysis the spectra of $\pi^{+}$ and $\pi^{-}$
at a given rapidity 
are known to be identical for transverse momenta from 170 MeV to 
about 800 MeV, the value up to which
$\pi^{+}$ can reliably be identified in the
experiment. 
Below $p_{t} =$ 170 MeV, the $\pi^{-}$ yield starts to exceed that of the 
$\pi^{+}$, which is attributed 
to the Coulomb effect. The results for the ratio  
$n(\Delta) / n({\mathrm{nucleon}} + \Delta)$ are summarized in the 
last column of Table~\ref{abund}. The number of nucleons excited to the \D  
resonance at freeze-out increases from 10 $\pm$ 2 $\%$ at 1.06 AGeV to 
18 $\pm$ 4 $\%$ at 1.93 AGeV. These values
are diplayed in Fig.~\ref{delapart} together with the result of 36 $\pm$ 5 $\%$
from a measurement at 14.6 AGeV\cite{e814-1}. Having  
determined the ratio $n(\Delta) / n({\mathrm{nucleon}} + \Delta)$, one can 
estimate the freeze-out temperature within the context of 
the hadrochemical equilibrium model\cite{pbm1}. The results are summarized 
in Table~\ref{ztT} together with those of the radial flow 
analysis\cite{bh1}; both analyses are in good agreement.

In summary, we have studied the pion phase space distribution in terms of
the thermal model including 
pions from the decay of \D resonances in addition to 
the thermal pions. The good agreement between the data and model 
gives confidence that the low-$p_{t}$ enhancement in the pion spectra 
is due to the pions from the \D decay.
When the beam energy is increased from 1.06 to 1.93 AGeV
this low-$p_{t}$ enhanced part of
the pion spectra  
decreases from 77 $\%$ to 66 $\%$ of the total pion multiplicity, while 
the number of nucleons excited to \D at freeze-out increases from 10 to 
18 $\%$ of all nucleons.
The freeze-out temperature which is estimated from the ratio 
$n(\Delta) / n({\mathrm{nucleon}} + \Delta)$
is found to be in good agreement with the results of the radial flow analyses.
The direct reconstruction of the \D by an invariant mass analysis 
of  ($\pi p$)- pairs is underway\cite{es1}, which will allow to compare
the results of the two different analyses in the future.\\

We would like to thank Prof. P. Braun-Munzinger for many helpful 
discussions about the thermal model and a careful reading of the manuscript.
This work was supported in part by the Bundesministerium
f\"{u}r Forschung und Technologie under the contracts 06 HD 525 I(3),
06 DR 666 I(3), X051.25/RUM-005-95 and X081.25/N-119-95.

\begin{table}
\caption{
Summary of results
for Ni + Ni collisions at the various incident energies (centrality
cut PM100): Average number of $\pi^{-}$ per event (2nd column), fraction
of the low-$\pi_t$ pion component (3rd column) and derived relative number of
nucleons excited to \D at freeze-out (last column).}
\vspace{0.5cm}
\begin{tabular}{cccc} 
$E_{beam}/A \mathrm{(GeV)}$ & 
$n(\pi^{-})$ & $\pi^{-}_{Low} / \pi^{-}_{All} (\%)$ &
$n(\Delta) / n({\mathrm{nucleon}} + \Delta) (\%)$ \\ \hline
1.06 & 3.6 $\pm$ 0.4 & 77 $\pm$ 4 & 10 $\pm$ 2 \\
1.45 & 5.8 $\pm$ 0.6 & 68 $\pm$ 4 & 13 $\pm$ 3 \\
1.93 & 8.5 $\pm$ 0.9 & 66 $\pm$ 3 & 18 $\pm$ 4 \\
\end{tabular}
\label{abund}
\end{table}
\vspace{1.0cm}

\begin{table}
\caption{
Comparison of $T$ in MeV obtained in this work ($\Delta$) and in a
radial flow } analysis~\cite{bh1}.
\vspace{0.5cm}
\begin{tabular}{ccc} 
$E_{beam}/A \mathrm{(GeV)}$ & \D & Flow \\ \hline
1.06 & 75 $\pm$ 5 & 79 $\pm$ 10 \\
1.45 & 80 $\pm$ 7 & 84 $\pm$ 10 \\
1.93 & 89 $\pm$ 9 & 92 $\pm$ 12 \\
\end{tabular}
\label{ztT}
\end{table}
\vspace{1.0cm}

\newpage

\newpage

\end{document}